# Temporal Optical Solitons Due to Kerr and Quintic Nonlinearities in Coupled Quantum Wells

S. Shwetanshumala[1,2], S. Konar[1*] and Anjan Biswas[3]

[1]Department of Applied Physics, Birla Institute of Technology, Mesra, Ranchi, Jharkhand, India
[2]Department of Physics, A.N. College, Patna, Bihar, India
[3]Department of Mathematical Sciences, Delaware State University, Dover, DE 19901-2277, USA

*Corresponding author: swakonar@yahoo.com

## Abstract

The possibility of generation and propagation of ultraslow bright optical solitons in asymmetric three-coupled quantum well systems is studied in this paper. These bright solitons owe their existence to Kerr and quintic nonlinearities which arise due to a probe pulse and two controlling laser beams. We also demonstrate numerically that these solitons are stable. The modulation instability of a continuous or quasi-continuous wave probe beam has been also investigated and the role of quintic nonlinearity in suppressing this instability addressed.

PACS number (s): 42.65.Tg, 42.50.Gy, 78.67.De

# I. INTRODUCTION

In recent years, optical solitons have received tremendous attention due to their potential applications in communication, information processing and optical computing [1-4]. In particular, keen interest has been shown in the studies on sub and superluminal optical soliton propagation due to quantum optical coherence and interference effects in semiconductor quantum wells (SQWs) [5-9]. This keen interest on optical solitons in quantum well (QW) structures is also owing to the fact that optical transitions between electronic states in QWs paving the way for new devices and their applications in communication and information processes [10]. Due to small effective electron mass, SQWs possess large electric dipole moments of inter subband transitions and high nonlinear optical coefficients. Furthermore, their transition energies, dipole moments and symmetries can be engineered as desired by choosing the materials and structure dimensions in device design. Based on quantum coherence and interference effects, Kerr nonlinearity, which is primarily responsible for creation of optical solitons, can be effectively enhanced while linear absorption and even two photon absorption can be well suppressed. In SQWs, optical solitons can be created at a very low power, of the order of a few mW.

The formation of optical solitons in QWs is the manifestation of a delicate balance between group velocity dispersion (GVD) and self phase modulation due to optical nonlinearity induced by the probe pulse and control fields. To date, only the role of Kerr nonlinearity has been considered to investigate soliton propagation in semiconductor quantum well nanostructures [5-9]. However, optical solitons in QW structures have much shorter time length and thus nonlinear effects are much stronger. Hence, it is necessary to examine whether higher order nonlinearities are also responsible for soliton formation in these systems. Thus, in this communication we show the existence of bright solitons based on bound–to-bound inter subband transition (ISBT) in an asymmetric three-coupled quantum well (TCQW) structure and investigate their dynamics when the system provides both Kerr and quintic nonlinearities. We have also shown that a continuous or quasi-continuous probe wave can undergo modulation instability.

The paper is organized as follows. In sec II, we present theoretical model and derive expression for optical nonlinearity. In this section we also derive nonlinear Schrödinger equation governing the motion of the envelope of the probe field. In sec III, we discuss formation of solitons. In sec IV, we investigate modulation instability of a continuous wave. Finally, section V contains a brief conclusion.

## II. MODEL AND EQUATIONS OF MOTION

To start with, we consider an asymmetric three-coupled quantum well (TCQW) structures having four electronic energy levels that forms the well known cascade configuration. The TCQW sample consists of 40 coupled well periods. Each well period consists of three GaInAs wells of thickness 4.2, 2.0 and 1.8 nm respectively and they are separated by 1.6 nm barriers made of AlInAs. The TCQW structure has a couple of advantages: (i) The effective mass of electrons in the barrier material (AlInAs) is significantly smaller than that in the high Al concentration compositions of AlGaAs. The latter one is required to confine four equally spaced bound states, so that the coupling barrier need not be problematically thin ( <10 Å). (ii) The small value of electron effective mass in GaInAs has the advantage of larger dipole matrix elements for inter subband transitions.

The present structure can have energy levels as $\varepsilon_1 = 151\,\text{meV}$, $\varepsilon_2 = 270\,\text{meV}$, $\varepsilon_3 = 386\,\text{meV}$ and $\varepsilon_4 = 506\,\text{meV}$. The energy levels of this system are shown in figure (1). Such structures have been already investigated by several authors for optical properties including higher order dispersion [9-11]. All possible transitions between subbands in the above quantum well are dipole allowed, where $\omega_{21}$, $\omega_{32}$, and $\omega_{43}$ respectively represent the energy difference of $|2\rangle \to |1\rangle, |3\rangle \to |2\rangle$ and $|4\rangle \to |3\rangle$ transitions. In the present analysis, we assume that the QW is designed with low doping such that electron-electron effects have very small influence in our study. As a result, many body effects arising out of electron-electron interactions are not important. A weak probe optical pulse with angular frequency $\omega_P$, wave vector $k_P = \omega_P / c$, polarization vector $\hat{e}_P$ and amplitude $E_P$ is assumed to propagate in $z$-direction inside the QW where it interacts with this four level system. The growth direction of the quantum well is also along the y-axis and z-axis is parallel to the QW plane. The quantum well system also interacts with two continuous wave (CW) control laser fields. The probe and coupling electromagnetic fields possess transverse polarization. The electric field associated with the probe and control fields can be written as

$$\vec{E}(z,t) = \hat{e}_P E_P \exp\left[i(k_P z - \omega_P t)\right] + \hat{e}_{C1} E_{C1} \exp\left[i(k_{C1} z - \omega_{C1} t)\right] + \hat{e}_{C2} E_{C2} \exp\left[i(k_{C2} z - \omega_{C2} t)\right] + c.c., \quad (1)$$

where $\hat{e}_{C1}$ and $\hat{e}_{C2}$ are the unit vectors along the polarization direction of the space-time independent strong coupling fields $E_{C1}$ and $E_{C2}$ respectively: $k_j$ and $\omega_j$ are respectively the wave vector and the angular frequency of the field whose slowly varying envelope is $E_j$, $j = C1$ and $C2$. Two different formalisms are available in literature to describe the system dynamics comprising quantum well and electromagnetic fields. The first one is the Schrodinger formalism which is also known as probability amplitude approach, where the state is described by a state function satisfying the Schrödinger equation. The other one is the density operator formalism, where the state is described by a density operator satisfying the master equation. In several recent investigations, it has been shown that the Schrödinger formalism is equivalent to the density operator formalism in describing the phenomena of electromagnetically induced transparency (EIT), EIT related four wave mixing and soliton propagation in quantum well, quantum dot and atomic systems [12-19]. In a recent publication [19], Jianbing Qi has studied EIT in an inverted Y-type four level system. In this study the author has employed both probability amplitude and density matrix methods to obtain optical susceptibility of the medium. Both the methods yield identical expression for susceptibility. However, since the probability amplitude formalism is much simpler than the density operator formalism, we adopt the former to describe the system consisting of quantum well and electromagnetic fields.

In the Schrödinger picture, the semi-classical Hamiltonian describing the system of electrons and electromagnetic fields can be written as a sum of two terms $\hat{H} = \hat{H}_0 + \hat{H}'$, where $\hat{H}_0$ describes the free Hamiltonian of the system in the absence of electromagnetic field and $\hat{H}'$ describes the interaction between quantum well electrons and applied field. In the Schrödinger picture, two parts of the Hamiltonian can be written respectively as

$$\hat{H}_0 = \sum_{j=1}^{4} \varepsilon_j \mid j \rangle \langle j \mid, \tag{2}$$

and

$$\hat{H}' = -\hbar \{ \Omega_P \exp[i(k_P z - \omega_P t)] |2\rangle\langle 1| + \Omega_{C1} \exp[i(k_{C1} z - \omega_{C1} t)] |3\rangle\langle 2|$$
$$+ \Omega_{C2} \exp[i(k_{C2} z - \omega_{C2} t)] |4\rangle\langle 3| + h.c. \} \quad , \quad (3)$$

where $\Omega_p$, $\Omega_{C1}$ and $\Omega_{C2}$ are the half Rabi frequencies corresponding to the laser driven inter subband transitions $|1\rangle \leftrightarrow |2\rangle$, $|2\rangle \leftrightarrow |3\rangle$ and $|3\rangle \leftrightarrow |4\rangle$ respectively. Half Rabi frequencies of the probe and control fields are related to the respective slowly varying amplitudes $E_p$, $E_{C1}$ and $E_{C2}$ through $\Omega_p = \frac{(\hat{\mu}_{12} \cdot \hat{e}_p) E_p}{2\hbar}$, $\Omega_{C1} = \frac{(\hat{\mu}_{23} \cdot \hat{e}_{C1}) E_{C1}}{2\hbar}$ and $\Omega_{C2} = \frac{(\hat{\mu}_{34} \cdot \hat{e}_{C2}) E_{C2}}{2\hbar}$, where $\hat{\mu}_{kl}$ being the dipole moment for the transition between subbands $|k\rangle$ and $|l\rangle$. It should be noted that another commonly used definition of half Rabi frequency is: $\Omega = \frac{|\mu|E}{\hbar}$, which differs by a factor of two from the definition used in present paper.

The state vector $|\Psi(t)\rangle_s$ of the system in the Schrödinger picture is

$$|\Psi(t)\rangle_s = \sum_{j=1}^{4} C_j(t) |j\rangle, \quad (4)$$

where $|j\rangle$ is the eigenstate of $\hat{H}_0$ and $C_j(t)$ is the normalized time dependent probability amplitude of finding the electron in the subband $|j\rangle$. Time evolution of the present system can be studied more conveniently in the interaction picture, and in order to do that we need to transform both $\hat{H}'$ and $|\Psi(t)\rangle_s$ in the interaction picture using the operator $\hat{U} = \exp(i\hat{H}_0 t/\hbar)$. In the interaction picture, the Hamiltonian and the state vector can be obtained using $\hat{H}_{int} = \hat{U}\hat{H}'\hat{U}^{-1}$ and $|\Psi(t)\rangle_{int} = \hat{U}|\Psi(t)\rangle_s$. Making use of $\hat{H}_0$ and the closure theorem [20] $\sum_j |j\rangle\langle j| = 1$, the operator $\hat{U}$ can be written as

$$\hat{U} = \exp(i\varepsilon_1 t/\hbar)|1\rangle\langle 1| + \exp(i\varepsilon_2 t/\hbar)|2\rangle\langle 2| + \exp(i\varepsilon_3 t/\hbar)|3\rangle\langle 3| + \exp(i\varepsilon_4 t/\hbar)|4\rangle\langle 4|. \qquad (5)$$

Applying $\hat{U}$ on $|\Psi(t)\rangle_s$, the interaction picture wave function $|\Psi(t)\rangle_{int}$ is readily obtained as

$$\begin{aligned}|\Psi(t)\rangle_{int} =\ & C_1(t)\exp(i\varepsilon_1 t/\hbar)|1\rangle + C_2(t)\exp(i\varepsilon_2 t/\hbar)|2\rangle + C_3(t)\exp(i\varepsilon_3 t/\hbar)|3\rangle \\ & + C_4(t)\exp(i\varepsilon_4 t/\hbar)|4\rangle.\end{aligned} \qquad (6)$$

It is straight forward to obtain the Hamiltonian in the interaction picture which turns out to be

$$\begin{aligned}\hat{H}_{int} = -\hbar\big\{ & \Omega_P \exp[i(k_P z - \omega_P t)]\exp[-i(\varepsilon_1 - \varepsilon_2)t/\hbar]|2\rangle\langle 1| + \Omega_{C1}\exp[i(k_{C1}z - \omega_{C1}t)] \\ & \exp[-i(\varepsilon_2 - \varepsilon_3)t/\hbar]|3\rangle\langle 2| + \Omega_{C2}\exp[i(k_{C2}z - \omega_{C2}t)]\exp[-i(\varepsilon_3 - \varepsilon_4)t/\hbar]|4\rangle\langle 3| + h.c.\big\}.\end{aligned} \qquad (7)$$

We are now in a position to use the Schrödinger equation $i\hbar\dfrac{\partial|\Psi(t)\rangle_{int}}{\partial t} = \hat{H}_{int}|\Psi(t)\rangle_{int}$ in the interaction picture to obtain the following equations of motion for probability amplitudes:

$$i\hbar\frac{\partial C_1}{\partial t} - \varepsilon_1 C_1 = -\hbar\Omega_P^* C_2 \exp\{-i(k_P z - \omega_P t)\}, \qquad (8a)$$

$$i\hbar\frac{\partial C_2}{\partial t} - \varepsilon_2 C_2 = -\hbar\Omega_P C_1 \exp\{i(k_P z - \omega_P t)\} - \hbar\Omega_{C1}^* C_3 \exp\{-i(k_{C1}z - \omega_{C1}t)\}, \qquad (8b)$$

$$i\hbar\frac{\partial C_3}{\partial t} - \varepsilon_3 C_3 = -\hbar\Omega_{C1} C_2 \exp\{i(k_{C1}z - \omega_{C1}t)\} - \hbar\Omega_{C2}^* C_4 \exp\{-i(k_{C2}z - \omega_{C2}t)\}, \qquad (8c)$$

$$i\hbar\frac{\partial C_4}{\partial t} - \varepsilon_4 C_4 = -\hbar\Omega_{C2} C_3 \exp\{i(k_{C2}z - \omega_{C2}t)\}. \qquad (8d)$$

Introducing the transformation $C_j(t) = a_j(t)\exp\left\{i\left[\vec{k}_j \cdot \vec{r} - \left(\frac{\varepsilon_j}{\hbar} + \theta_j\right)t\right]\right\}$, with $\vec{k}_1 = 0$, $\vec{k}_2 = k_P\hat{e}_z$, $\vec{k}_3 = (k_P + k_{c1})\hat{e}_z$, $\vec{k}_4 = (k_P + k_{c1} + k_{c2})\hat{e}_z$, $\theta_1 = 0$, $\theta_2 = \Delta_1$, $\theta_3 = \Delta_2$ and $\theta_4 = \Delta_3$, the probability amplitude equations (8a) to (8d) reduce to the following form:

$$\frac{\partial a_1}{\partial t} = i\Omega_p^* a_2,\tag{9a}$$

$$\frac{\partial a_2}{\partial t} = i\Delta_1 a_2 + i\Omega_p a_1 + i\Omega_{C1}^* a_3,\tag{9b}$$

$$\frac{\partial a_3}{\partial t} = i\Delta_2 a_3 + i\Omega_{C1} a_2 + i\Omega_{C2}^* a_4,\tag{9c}$$

$$\frac{\partial a_4}{\partial t} = i\Delta_3 a_4 + i\Omega_{C2} a_3,\tag{9d}$$

where the detunings $\Delta_1$, $\Delta_2$ and $\Delta_3$ are defined as $\Delta_1 = \omega_P - \frac{\varepsilon_2 - \varepsilon_1}{\hbar}$, $\Delta_2 = \omega_{C1} - \frac{\varepsilon_3 - \varepsilon_2}{\hbar} + \Delta_1$ and $\Delta_3 = \omega_{C2} - \frac{\varepsilon_4 - \varepsilon_3}{\hbar} + \Delta_2$. At this stage we introduce decay rates $\gamma_j (j = 2, 3, 4)$ phenomenologically to describe the corresponding total decay rate of the subband $|j\rangle$. A comprehensive treatment of the decay rates would involve incorporation of the decay mechanisms into the Hamiltonian of the system. However, decay rates have been widely used phenomenologically under similar situation, and are very successful in estimating optical nonlinearity in quantum well systems. Therefore, we also introduce decay rates phenomenologically in the present investigation just as it is done in the literature [7, 21-24]. It should be mentioned that in semiconductor QW, total decay rate $\gamma_j$ of the subband $|j\rangle$ consists of population decay rate $\gamma_{jl}$ which is primarily due to longitudinal optical phonon emission events at low temperature, and dephasing rates $\gamma_j^{dph}$ which depends on electron-electron scattering, electron-phonon scattering and on inhomogeneous broadening due to scattering on interface roughness. After the introduction of decay rates, equations (9a)- (9d) take the following form

$$\frac{\partial a_1}{\partial t} = i\Omega_p^* a_2, \tag{10a}$$

$$\frac{\partial a_2}{\partial t} = i\Delta_1 a_2 + i\Omega_p a_1 + i\Omega_{C1}^* a_3 - \gamma_2 a_2, \tag{10b}$$

$$\frac{\partial a_3}{\partial t} = i\Delta_2 a_3 + i\Omega_{C1} a_2 + i\Omega_{C2}^* a_4 - \gamma_3 a_3, \tag{10c}$$

$$\frac{\partial a_4}{\partial t} = i\Delta_3 a_4 + i\Omega_{C2} a_3 - \gamma_4 a_4. \tag{10d}$$

The evolution of the electric field of the probe pulse is governed by the Maxwell equation:

$$\nabla^2 \vec{E} - \frac{1}{c^2}\frac{\partial^2 \vec{E}}{\partial t^2} = \frac{1}{\varepsilon_0 c^2}\frac{\partial^2 \langle\vec{P}\rangle}{\partial t^2}, \tag{11}$$

where $\langle\vec{P}\rangle$ denotes the polarization induced in the medium by the probe and control fields and is given by

$$\langle\vec{P}\rangle = N \langle\vec{d}\rangle, \tag{12}$$

$N$ is the electron density in the quantum well and $\langle \vec{d} \rangle$ is the dipole moment operator. For simplicity, we assume that the probe field is homogeneous in transverse direction. Therefore, under slowly varying envelope approximation [25], the amplitude of the probe laser field $E_P = E_P(z,t)$ propagating along $z$-direction obeys:

$$\frac{\partial \Omega_P}{\partial z} + \frac{1}{c}\frac{\partial \Omega_P}{\partial t} = i\kappa a_2 a_1^*, \tag{13}$$

where $\kappa = \frac{N|\mu_{12}|^2 \omega_P}{2\hbar\varepsilon_0 c}$. We now assume that the system is initially in the ground state $|1\rangle$ i.e., $a_1 = 1$, $a_2 = a_3 = a_4 = 0$ at $t = 0$. Notice that the probe field $\Omega_P$ is much weaker than that of the coupling fields $\Omega_{C1}$ and $\Omega_{C2}$ and hence due to quantum coherence and interference effects, the depletion of the ground state is not significant and thus $a_1 \simeq 1$ for $t > 0$. This is a usual approximation in this area which has been used frequently and successfully to obtain optical nonlinearities in semiconductor QW nanostructures [26-30]. This assumption has been also used to identify giant Kerr nonlinearities in coupled double quantum well nanostructures [30] and also in a crystal of molecular magnets [14]. Hence, we also adopt this approximation in our present investigation. We assume that $a_j = \sum_k a_j^{(k)}$, $a_j^{(k)}$ is the $k$th order part of $a_j$ in terms of $\Omega_P$. With the adiabatic framework as outlined above, it can be shown that $a_j^{(0)} = \delta_{j1}$ and $a_1^{(1)} = 0$, where $\delta_{j1}$ is the Kronecker delta. In view of above assumption, equation (13) reduces to

$$\frac{\partial \Omega_P}{\partial z} + \frac{1}{c}\frac{\partial \Omega_P}{\partial t} = i\kappa a_2^{(1)} + (NLT)a_2^{(1)}, \tag{14}$$

where $NLT = -i\kappa\left[|a_2^{(1)}|^2 + |a_3^{(1)}|^2 + |a_4^{(1)}|^2 - \left(|a_2^{(1)}|^2 + |a_3^{(1)}|^2 + |a_4^{(1)}|^2\right)^2\right]$.

Taking the time Fourier transform of equations (10a) - (10d) and keeping terms up to first order of $\Omega_p$, we have

$$(\omega+\Delta_1+i\gamma_2)\alpha_2^{(1)}+\Omega_{C1}^*\alpha_3^{(1)}+\Lambda_P=0, \tag{15a}$$

$$(\omega+\Delta_2+i\gamma_3)\alpha_3^{(1)}+\Omega_{C1}\alpha_2^{(1)}+\Omega_{C2}^*\alpha_4^{(1)}=0, \tag{15b}$$

$$(\omega+\Delta_3+i\gamma_4)\alpha_4^{(1)}+\Omega_{C2}\alpha_3^{(1)}=0, \tag{15c}$$

where $\alpha_j^{(1)}$ and $\Lambda_P$ are the Fourier transform of $a_j^{(1)}$ and $\Omega_p$ respectively and $\omega$ is the Fourier transform variable. Taking Fourier transform of the linearized version of equation (14), we obtain:

$$\frac{\partial \Lambda_P}{\partial z}-i\beta(\omega)\Lambda_P=0, \tag{16}$$

where $\beta(\omega)=\frac{\omega}{c}-\kappa\frac{D_P(\omega)}{D(\omega)}$, $D_P(\omega)=(\omega+\Delta_2+i\gamma_3)(\omega+\Delta_3+i\gamma_4)-|\Omega_{C2}|^2$, and

$$D(\omega)=(\omega+\Delta_1+i\gamma_2)(\omega+\Delta_2+i\gamma_3)(\omega+\Delta_3+i\gamma_4)-|\Omega_{C2}|^2(\omega+\Delta_1+i\gamma_2)$$
$$-|\Omega_{C1}|^2(\omega+\Delta_3+i\gamma_4).$$

The term $\beta(\omega)$ can be identified as the frequency dependent dispersion function. Equation (16) can be solved analytically to obtain

$$\Lambda_P(z,\omega)=\Lambda_P(0,\omega)\exp[i\beta(\omega)z]. \tag{17}$$

To investigate properties of the propagating probe pulse, the propagation constant $\beta = \beta(\omega)$ can be expanded in Taylor series around the central frequency of the probe field i.e. $\omega = 0$, as

$$\beta(\omega) = \beta(0) + \beta'(0)\omega + \frac{1}{2}\beta''(0)\omega^2 + 0(\omega^3), \quad (18)$$

$$\text{where } \beta(0) = -\kappa \frac{D_p(0)}{D(0)}, \quad (19)$$

$$\beta'(0) = \frac{1}{c} - \kappa \frac{(\Delta_2 + \Delta_3 + i\gamma_3 + i\gamma_4)}{D(0)}$$
$$+ \kappa \frac{D_P(0)\left[D_P(0) + (\Delta_1 + i\gamma_2)(\Delta_3 + i\gamma_4) + (\Delta_1 + i\gamma_2)(\Delta_2 + i\gamma_3) - |\Omega_{C1}|^2\right]}{[D(0)]^2}, \quad (20)$$

$$\beta''(0) = -\frac{2\kappa}{D(0)}$$

$$+ \frac{2\kappa(\Delta_2 + \Delta_3 + i\gamma_3 + i\gamma_4)\{D_P(0) + (\Delta_1 + i\gamma_2)(\Delta_3 + i\gamma_4) + (\Delta_1 + i\gamma_2)(\Delta_2 + i\gamma_3) - |\Omega_{C1}|^2\}}{[D(0)]^2}$$

$$- \frac{2\kappa D_P(0)\{D_P(0) + (\Delta_1 + i\gamma_2)(\Delta_3 + i\gamma_4) + (\Delta_1 + i\gamma_2)(\Delta_2 + i\gamma_3) - |\Omega_{C1}|^2\}^2}{[D(0)]^3}$$

$$+ \frac{2\kappa D_P(0)\{\Delta_1 + \Delta_2 + \Delta_3 + i\gamma_2 + i\gamma_3 + i\gamma_4\}}{[D(0)]^2}, \quad (21)$$

$$D_P(0) = (\Delta_2 + i\gamma_3)(\Delta_3 + i\gamma_4) - |\Omega_{C2}|^2, \quad (22)$$

$$D(0) = (\Delta_1 + i\gamma_2)(\Delta_2 + i\gamma_3)(\Delta_3 + i\gamma_4) - |\Omega_{C2}|^2(\Delta_1 + i\gamma_2) - |\Omega_{C1}|^2(\Delta_3 + i\gamma_4). \quad (23)$$

Physically, $\beta'(0)$ is related to the group velocity $v_g$ of the probe field $\left(v_g = \text{Re}\left(\frac{1}{\beta'(0)}\right)\right)$ and real part of $\beta''(0)$ represents group velocity dispersion which is responsible for broadening of laser probe pulse. Under specific condition, $\beta''(0)$ together with optical nonlinearity are responsible for creation of optical solitons which travel with a group velocity $v_g$. Real part of $\beta(0)$ describes phase shift of the probe field, whereas imaginary part signifies linear absorption. Imaginary part of $\beta''(0)$ determines nonlinear loss/ gain of the probe field. Please note that, it is possible to find out suitable parameter regime in which the absorption of the probe field can be substantially reduced or enhanced due to interference produced by the two control fields.

Equation (16) is obtained using linearized wave equation where optical nonlinearity of the probe field has been neglected. To investigate nonlinear pulse propagation we need to incorporate the effect of optical nonlinear terms in the pulse dynamics. These nonlinear terms are responsible for self phase modulation phenomenon which together with GVD leads to shape preserving solitary wave propagation. Therefore, to proceed further, instead of considering equation (16), we take equation (14) which can be put in the following form

$$\frac{\partial \Lambda_P}{\partial z} = i\beta(\omega)\Lambda_P + (NLT)\alpha_2^{(1)}. \tag{24}$$

Please note that equation (16) is recovered from equation (24) by ignoring the nonlinear part. We follow the method developed by Wu and Deng [30,31], take a trial function $\Lambda_P(z,\omega) = \tilde{\Lambda}_P(z,\omega)\exp[i\beta(0)z]$, submit it in the above equation and keep terms up to order $\omega^2$ in $\beta(\omega)$ to obtain

$$\left[\frac{\partial}{\partial z}-i\omega\beta'(0)-i\frac{\omega^2}{2}\beta''(0)\right]\tilde{\Lambda}_P \exp\left[i\beta(0)z\right]=(NLT)\alpha_2^{(1)}. \qquad (25)$$

Performing the inverse Fourier transform $\tilde{\Omega}_P(z,t)=\frac{1}{\sqrt{2\pi}}\int_{-\infty}^{+\infty}\tilde{\Lambda}_P(z,\omega)\exp(-i\omega t)d\omega$, we get,

$$\left(\frac{\partial \tilde{\Omega}_P}{\partial z}+\beta'(0)\frac{\partial \tilde{\Omega}_P}{\partial t}+\frac{i}{2}\beta''(0)\frac{\partial^2 \tilde{\Omega}_P}{\partial t^2}\right)\exp\{i\beta(0)z\}=(NLT)a_2^{(1)}. \qquad (26)$$

Inserting the value of $|a_2^{(1)}|^2$, $|a_3^{(1)}|^2$ and $|a_4^{(1)}|^2$ in the $NLT$, we immediately obtain

$$i\frac{\partial \tilde{\Omega}_P}{\partial z}+i\beta'(0)\frac{\partial \tilde{\Omega}_P}{\partial t}-\frac{1}{2}\beta''(0)\frac{\partial^2 \tilde{\Omega}_P}{\partial t^2}+W\exp(-\alpha z)|\tilde{\Omega}_P|^2\tilde{\Omega}_P$$
$$-M\exp(-2\alpha z)|\tilde{\Omega}_P|^4\tilde{\Omega}_P=0, \qquad (27)$$

where $W=\kappa\dfrac{D_P(0)}{D(0)}\left\{\dfrac{|D_P(0)|^2+|\Omega_{C1}|^2\left[|\Delta_3+i\gamma_4|^2+|\Omega_{C2}|^2\right]}{|D(0)|^2}\right\}$,

$M=\kappa\dfrac{D_P(0)}{D(0)}\left\{\dfrac{|D_P(0)|^2+|\Omega_{C1}|^2\left[|\Delta_3+i\gamma_4|^2+|\Omega_{C2}|^2\right]}{|D(0)|^2}\right\}^2$,

and $\alpha=2\operatorname{Im}[\beta(0)]$.

## III. SOLITONS DUE TO KERR AND QUINTIC NONLINEARITY

Introducing the retarded frame defined by $\xi=z$ and $\eta=t-z\beta'(0)$, equation (27) can be recasted as

$$i\frac{\partial \tilde{\Omega}_P}{\partial \xi}-\frac{1}{2}\beta''(0)\frac{\partial^2 \tilde{\Omega}_P}{\partial \eta^2}+W\exp(-\alpha z)|\tilde{\Omega}_P|^2\tilde{\Omega}_P-M\exp(-2\alpha z)|\tilde{\Omega}_P|^4\tilde{\Omega}_P=0. \qquad (28)$$

In general, coefficients $W$ and $M$ are complex and therefore, equation (28) does not possess soliton solution. However, for suitable set of system parameters, imaginary parts of these co-efficients may be made very small in comparison to their real parts. Under such situations, $\beta'' = \beta_r'' + i\beta_i'' \simeq \beta_r''$, $W = W_r + iW_i \simeq W_r$ and $M = M_r + iM_i \simeq M_r$. Therefore, it is possible to obtain shape preserving soliton solution which propagates over long distance without distortion. Equation (28) is modified nonlinear Schrödinger equation (MNLSE) with quintic nonlinearity. In the absence of the quintic term and when the absorption co-efficient $\alpha$ is small, equation (28) is the standard nonlinear Schrödinger equation which admits solutions describing bright and dark solitons. Several authors have earlier studied soliton propagation in QW systems [4-9] with Kerr type nonlinearity (i.e., $M = 0$). The sign of the product $\text{Re}\left[\beta''(0)\right].\text{Re}(W)$ determines the nature of the soliton. For bright solitons $\text{Re}\left[\beta''(0)\right].\text{Re}(W) < 0$, while for dark solitons $\text{Re}\left[\beta''(0)\right].\text{Re}(W) > 0$. Finite value of $M$ introduces higher order nonlinearity in the system which gives rise to existence of entirely different types of soliton solution including bistable one. We now present numerical example to demonstrate at first the existence of Kerr optical solitons and subsequently solitons due to Kerr and quintic nonlinearities. The parameters taken in our study are as follows: $N = 10^{16} \, cm^{-3}$, $\mu_{12} = 13 \, eA^0$, $\omega_p = 18.08 \times 10^{13} \, s^{-1}$, thus $\kappa = 1.4 \times 10^{11} \, \mu m^{-1} s^{-1}$, decay rates $\gamma_2 = 4.8 \times 10^9 \, s^{-1}$, $\gamma_3 = 3.4 \times 10^9 \, s^{-1}$, $\gamma_4 = 4.2 \times 10^{11} \, s^{-1}$, Rabi frequencies $\Omega_{C1} = 3.5 \times 10^{11} \, s^{-1}$, $\Omega_{C2} = 4.8 \times 10^{11} \, s^{-1}$, detunings $\Delta_1 = -1.0 \times 10^{11} \, s^{-1}$, $\Delta_2 = -2.0 \times 10^{12} \, s^{-1}$ and $\Delta_3 = -4.0 \times 10^{12} \, s^{-1}$. With these parameters, $\beta(0) = 3.78 + i0.12 \, \mu m^{-1}$, $\beta''(0) = 5.88 \times 10^{-21} + i5.91 \times 10^{-22} \, \mu m^{-1} s^2$, $W = -2.85 \times 10^{-21} - i9.48 \times 10^{-23} \, \mu m^{-1} s^2$, $M = -2.16 \times 10^{-42} - i7.16 \times 10^{-44} \, \mu m^{-1} s^4$, $\alpha = 2 \times \text{Im}\left[\beta(0)\right] = 0.24 \, \mu m^{-1}$. In the present example $|W_i| \ll |W_r|$, $|M_i| \ll |M_r|$ and $|\beta_i''(0)| \ll |\beta_r''(0)|$, hence, equation (28) would admit soliton solutions. In order to investigate soliton solutions of equation (28), we first recast it, ignoring imaginary parts of $W$, $M$ and $\beta''(0)$, in the following form:

$$i\frac{\partial \tilde{\Omega}_P}{\partial Z} - \frac{s_D}{2}\frac{\partial^2 \tilde{\Omega}_P}{\partial \tau^2} + s_K |\tilde{\Omega}_P|^2 \tilde{\Omega}_P - s_Q \delta |\tilde{\Omega}_P|^4 \tilde{\Omega}_P = 0, \qquad (29)$$

where $s_D$, $s_K$ and $s_Q$ represent sign of real part of $\beta''$, $W$ and $M$, respectively; $Z = |W_r|\xi$, $\tau = \left[|W_r|/|\beta_r''(0)|\right]^{1/2}\eta$ and $\delta = |M_r|/|W_r|$.

In the present case both $s_K$ and $s_Q$ are negative, hence, signs of the Kerr (third term in equation(29)) and quintic (fourth term in equation(29)) nonlinearities are opposite. In general, almost all natural systems exhibiting optical nonlinearity show saturating form of nonlinearity, where, the nonlinearity increases linearly with the intensity of light when it is weak and saturates to a constant value when intensity of light is large. If both cubic and quintic terms are of the same sign, then the magnitude of total nonlinearity of the system would rise faster in comparison to the contribution of Kerr nonlinearity alone. This would lead to a situation where the magnitude of nonlinearity of the system becomes unbounded. A negative (positive) sign for cubic i.e., Kerr term and a positive (negative) sign for quintic term ensure saturating nature of total optical nonlinearity of the system which is similar to the behavior of many other physical systems exhibiting saturating form of optical nonlinearity [1, 33-36].

In what follows we are going to prove that a combination of positive (negative) sign of $s_D$ and a negative (positive) sign of $s_K$ is essential for a bright soliton solution when the quintic term is ignored. We shall perform this task adopting variational formalism [37,38], which has been widely used in the optical soliton community to investigate self trapped optical pulse propagation under unmodified and modified nonlinear Schrödinger equations. It is a trivial task to show that equation (29) can be derived from the following Lagrangian $(L)$:

$$L = \frac{i}{2}\left(\tilde{\Omega}_P \frac{\partial \tilde{\Omega}_P^*}{\partial Z} - \tilde{\Omega}_P^* \frac{\partial \tilde{\Omega}_P}{\partial Z}\right) - \frac{s_D}{2}\left|\frac{\partial \tilde{\Omega}_P}{\partial \tau}\right|^2 - \frac{s_K}{2}|\tilde{\Omega}_P|^4 + \frac{s_Q \delta}{3}|\tilde{\Omega}_P|^6, \tag{30}$$

where the asterisk denotes the complex conjugate. For further investigation we use following trial function: $\tilde{\Omega}_P(Z,\tau) = A(Z)\mathrm{sec}\,h\{\tau/\sigma(Z)\}e^{i(\rho(Z)\tau^2 + \phi(Z))}$, (31) where $A(Z), \sigma(Z), \rho(Z)$ and $\phi(Z)$ are real amplitude, duration, chirp and longitudinal phase of the optical pulse, respectively. We can obtain a reduced Lagrangian $\langle L \rangle$ by inserting the trial function into the Lagrangian and integrating it over $\tau$ from $-\infty$ to $+\infty$. Thus,

$$\langle L \rangle = \int_{-\infty}^{+\infty} L\,d\tau$$

$$= A^2\left(\frac{\pi^2 \sigma^3}{6}\frac{d\rho}{dZ} + 2\sigma\frac{d\phi}{dZ}\right) - s_D A^2\left(\frac{1}{3\sigma} + \frac{\pi^2 \rho^2 \sigma^3}{3}\right) - \frac{2}{3}s_K \sigma A^4 + \frac{16}{45}s_Q \delta \sigma A^6. \tag{32}$$

Employing Rayleigh-Ritz optimization procedure and after some algebra, we get following equations:

$$A^2(Z)\sigma(Z) = A^2(0)\sigma(0) = \text{constant} = E_0 \text{ (say)}, \tag{33}$$

$$\rho(Z) = -\frac{1}{2s_D}\frac{d}{dZ}\left(\ln(\sigma(Z))\right), \tag{34}$$

and
$$\frac{d^2\sigma(Z)}{dZ^2} = \frac{4s_D^2}{\pi^2\sigma^3(Z)} + \frac{4s_K s_D E_0}{\pi^2\sigma^2(Z)} - \frac{64 s_Q s_D \delta E_0^2}{15\pi^2\sigma^3(Z)} . \quad (35)$$

Equation (35) can be derived from a potential $V(\sigma)$, such that $\frac{d^2\sigma}{dZ^2} = -\frac{dV}{d\sigma}$, where

$$V(\sigma) = \frac{2s_D^2}{\pi^2\sigma^2} + \frac{4s_K s_D E_0}{\pi^2\sigma} - \frac{32 s_Q s_D \delta E_0^2}{15\pi^2\sigma^2} . \quad (36)$$

For stable self-trapped/soliton solutions of equation (29), the potential should possess a minimum. These self-trapped solitary wave solutions correspond to specific $\sigma$ and $E_0$ values such that $\frac{dV}{d\sigma} = 0$, and $\frac{d^2V}{d\sigma^2} \succ 0$. The value of $\sigma = \sigma_{cr}$ for which $\frac{dV}{d\sigma} = 0$ can be easily obtained as:

$$\sigma = \sigma_{cr} = \frac{\frac{16}{15}s_Q \delta E_0^2 - s_D}{s_K E_0} . \quad (37)$$

Therefore, an optical pulse with pulse width $\sigma = \sigma(0) = \sigma_{cr}$ and peak amplitude $A(0) = \sqrt{E_0/\sigma(0)} = \sqrt{E_0/\sigma_{cr}}$ will propagate without changing its shape. The stability of these solutions can be examined from the second derivative of $V(\sigma)$ evaluated at the point $\sigma = \sigma_{cr}$, which turns out to be:

$$\left.\frac{d^2V}{d\sigma^2}\right|_{\sigma=\sigma_{cr}} = -\frac{4s_D s_K E_0}{\pi^2\sigma_{cr}^3} . \quad (38)$$

From equation (37), it appears that if we neglect quintic term ( i.e., setting $\delta = 0$), then to obtain self trapped solutions of equation (29), we require $\sigma = \sigma_{cr} = \dfrac{-s_D}{s_K E_0}$. Since $\sigma$ is the width of the optical pulse, we should have $\sigma \succ 0$. Therefore, a combination of $s_D$ positive (negative) and $s_K$ negative (positive) is necessary for bright soliton solution. In the present case, $s_D$ is positive and $s_K$ is negative, hence, equation (29) will possess bright soliton solutions if we neglect the quintic term. At this stage it is worth pointing out that, in standard optical fibers above 1.3μm, solitons are obtained as a result of balance between anomalous dispersion and self focusing Kerr nonlinearity [2]. However, in the present case, fundamental bright soliton results as a balance between normal dispersion ( i.e., $\beta_r''(0)$ positive ) and defocusing Kerr nonlinearity. In fact defocusing Kerr nonlinearity is essential since dispersion is normal in the present case.

To discuss Kerr solitons, we ignore the quintic term (i.e., $\delta = 0$ ) temporarily in equation (29). For parameters in the present example we find $v_g / c = 3.14 \times 10^{-5}$, which shows the soliton propagates in the subluminal regime. For a pulse of duration $\tau_0$, we find characteristic dispersion length $L_D = \tau_0^2 / |\beta_r''(0)|$ and nonlinear length $L_{NL} = 1/(|\tilde{\Omega}_{P0}|^2 |W_r|)$, where $\tilde{\Omega}_{P0}$ gives the peak power of the soliton. The fundamental bright soliton solution of equation(29) ( when $\delta = 0$ ) has the form[2]:

$$\tilde{\Omega}_P(Z,\tau) = \tilde{\Omega}_{P0} \operatorname{sech}\left(\tau/\tau_0\right) \exp\left[-\dfrac{i}{2}|W_r||\tilde{\Omega}_{P0}|^2 Z\right], \qquad (39)$$

where $\tilde{\Omega}_{P0}$ and temporal width $\tau_0$ of the soliton are related through $\tilde{\Omega}_{P0} = \frac{1}{\tau_0}\left[\frac{|\beta_r''(0)|}{|W_r|}\right]^{1/2}$. For a

75 $ps$ soliton, $|E_P|_{max} = \frac{\hbar}{\mu_{12}}\frac{1}{\tau_0}\left[\frac{|\beta_r''(0)|}{|W_r|}\right]^{1/2} = 193.59 \text{Vcm}^{-1}$. Peak intensity of the soliton turns

out to be $I_{max} = 2\varepsilon_0 c n_P |E_P|_{max}^2 = 1.44 \times 10^3 W cm^{-2}$, where probe refractive index

$n_P = 1 + c\frac{\beta_r(0)}{\omega_P} = 7.27$. Peak power of soliton $P_0 = I_{max} S_0 = 452 mW$, where $S_0$ is the cross-

sectional area of the probe laser pulse and radius of the spot size of the probe laser pulse has been taken to be $100 \mu m$.

Thus far, we have confined our discussion ignoring the quintic term in equation (29). However, in the parameter regime considered in the present investigation, we find $\frac{|M_r||\tilde{\Omega}_P|^4}{|W_r||\tilde{\Omega}_P|^2} \sim 0.28$, hence, the quintic term is significant. Therefore, we need to consider soliton propagation incorporating quintic term, since, inevitably this will reveal new propagation properties. However, at this stage, we need to ensure that the presence of large quintic nonlinearity does not violate our basic assumption that the probe pulse is weak in comparison to two coupling laser beams. Using the values of $\Omega_{C1}$ and $\Omega_{C2}$, we can easily find the values of $E_{C1}$ and $E_{C2}$ which turn out to be 2028.85 $V/cm$ and 2794.73 $V/cm$, respectively, where we have used $\mu_{23} = 22.7\ eA^0$ and used $\mu_{34} = 22.6\ eA^0$ [39]. With these electric fields, the intensities $I_{C1}$ and $I_{C2}$ of those two controlling beams turn out to be $I_{C1} \approx 74.34 \times 10^3 W/cm^2$ and $I_{C2} \approx 14.10 \times 10^4 W/cm^2$. On the other hand, peak intensity $(I_p)$ of the probe soliton pulse $I_p = I_{max} = 1.44 \times 10^3 W/cm^2$. Therefore, since $I_p \ll I_{C1}$ and $I_p \ll I_{C2}$, it is amply clear that within the limit of weak probe pulse the quintic nonlinearity could be large, and hence, cannot be ignored.

Following an analysis identical with that of reference [38], we now examine the implications of the sign of quintic nonlinearity in equation (29) on the stability of self-trapped/solitary wave propagation. In the present case, $s_D$ is positive, $s_K$ and $s_Q$ are negative i.e., sign of quintic nonlinearity is positive, therefore, from equation (37) $\sigma_{cr}$ is always positive for any finite value of $E_0$. Moreover, from equation (38), $\left.\frac{d^2V}{d\sigma^2}\right|_{\sigma=\sigma_{cr}}$ is also positive for any finite value of $\sigma_{cr}$. Therefore, self trapped solutions are always stable when $s_Q$ is negative. On the other hand, if the sign of quintic nonlinearity is negative i.e., $s_Q$ is positive, then, $\sigma_{cr}$ and $\left.\frac{d^2V}{d\sigma^2}\right|_{\sigma=\sigma_{cr}}$ are both positive only when $E_0^2 \prec \frac{15}{16\delta}$. Therefore, stable self-trapped/solitary wave solution is admissible only when $E_0^2 \prec \frac{15}{16\delta}$.

We now proceed to identify soliton solution of equation (29) taking quintic nonlinearity into account ( i.e., $\delta \neq 0$) . In equation (29), if $s_D$ is negative, $s_K$ and $s_Q$ are positive, then in such cases this equation is known as nonlinear cubic-quintic Schrödinger equation, whose solitary wave solution and related stabilities are well known[40-44]. However, in the present case, $s_D$ is positive, $s_K$ and $s_Q$ are negative, yet the soliton solution is identical with those mentioned in above references [40-44]. The soliton like solution of equation (29) can be obtained for any arbitrary value of $\delta$, which may be written as [40,41,43]:

$$\tilde{\Omega}_P(Z,\tau) = \frac{2^{1/2}\Lambda\exp(-i\Lambda^2 Z/2)}{\left[1+\left(1-\frac{8}{3}\delta\Lambda^2\right)^{1/2}\cos h(2\Lambda\tau)\right]^{1/2}}, \tag{40}$$

where the parameter $\Lambda$ is related to soliton amplitude and width and thus it determines soliton energy. In the limit $\delta = 0$, above soliton like solution reduces to single soliton solution of nonlinear Schrödinger equation: $\tilde{\Omega}_P(Z,\tau) = \Lambda\sec h(\Lambda\tau)\exp[-i\Lambda^2 Z/2]$. Solitary wave solution of the form given by equation (40) possess bistable property [40,41] i.e., two different solitary waves have different amplitudes but same width (pulse duration). However, strictly speaking, unlike Kerr solitons, above solitary wave solution is not a soliton since collision between two such solitary wave solutions is inelastic [1, 44]. In fact, under specific situation, two solitary waves of above kind may coalesce to form a single solitary wave, a similar situation will never occur in case of Kerr solitons. In spite of this shortcoming, in the optical soliton community, these solitary wave solutions are also known as solitons.

The expression for soliton energy $Q$ can be obtained using

$$Q = \int_{-\infty}^{+\infty} |\tilde{\Omega}_P|^2 \, d\tau = -\left(\frac{6}{\delta}\right)^{1/2}\tan h^{-1}\left\{\frac{6^{1/2}\left[-3+\left(9-24\delta\Lambda^2\right)^{1/2}\right]}{12\Lambda\delta^{1/2}}\right\}. \tag{41}$$

To ensure that $Q$ is always positive, we must have $\delta\Lambda^2 < 3/8$. In order to verify the stability of the above soliton we have numerically solved equation (29) by employing split step Fourier method. In figure (2) we have demonstrated the results of soliton propagation using full numerical simulation of equation (29). We find that the soliton is robust and stable as it propagates in the quantum well.

## IV. MODULATION INSTABILITY OF A PROBE BEAM

We now proceed to investigate the modulation instability of a continuous wave (CW) probe beam. Considering a steady state continuous wave solution of the form $\tilde{\Omega}_P(\xi,\eta) = \left[r^{1/2} + b(\xi,\eta)\right]\exp\left\{i\left(W_r r - M_r r^2\right)\xi\right\}$ of equation (28), we obtain

$$i\frac{\partial b}{\partial \xi} - \frac{1}{2}\beta_r''(0)\frac{\partial^2 b}{\partial \eta^2} + \left(W_r r - 2M_r r^2\right)(b+b^*) = 0 , \quad (42)$$

where $r^{1/2}$ is the amplitude of CW and $b(\xi,\eta)$ is a small perturbation in amplitude over the CW and we have ignored linear attenuation. Assuming the perturbation $b(\xi,\eta)$ to be composed of two side band plane waves, we take modulation ansatz of the following form

$$b(\xi,\eta) = b_1(\xi)\exp(ip\eta) + b_2(\xi)\exp(-ip\eta), \quad (43)$$

where $b_1(\xi)$ and $b_2(\xi)$ are complex amplitudes of the perturbation field and $p$ is the angular frequency of the side band. Substituting the ansatz for the perturbation field into its evolution equation, we obtain a system of two coupled first order ordinary differential equations for the perturbation amplitudes of the sidebands:

$$i\frac{\partial b_1}{\partial \xi} + \frac{1}{2}\beta''(0)p^2 b_1 + \left(W_r r - 2M_r r^2\right)\left(b_1 + b_2^*\right) = 0 \quad (44a)$$

$$i\frac{\partial b_2}{\partial \xi} + \frac{1}{2}\beta''(0)p^2 b_2 + \left(W_r r - 2M_r r^2\right)\left(b_1^* + b_2\right) = 0 \quad (44b)$$

Above equations can be decoupled to obtain an equivalent set of two ordinary second order differential equations for $b_1$ and $b_2$ :

$$\frac{d^2 b_1}{d\xi^2} = \left[ \beta_r''(0) p^2 \left( |W_r| r - 2 |M_r| r^2 \right) - \frac{1}{4} \beta_r''(0)^2 p^4 \right] b_1 \tag{45}$$

$$\frac{d^2 b_2}{d\xi^2} = \left[ \beta_r''(0) p^2 \left( |W_r| r - 2 |M_r| r^2 \right) - \frac{1}{4} \beta_r''(0)^2 p^4 \right] b_2. \tag{46}$$

The perturbation field amplitudes $b_1$ and $b_2$ grow exponentially as long as the quantity in the bracket remains real and positive. The gain $g(p)$ of the instability turns out to be

$$g(p) = 2 \operatorname{Re} \left[ \beta_r''(0) p^2 \left( |W_r| r - 2 |M_r| r^2 \right) - \frac{1}{4} \beta_r''(0)^2 p^4 \right]^{1/2} \tag{47}$$

At the maximum of $g(p)$, the modulation instability sets in resulting in the breakup of the CW beam into a periodic pulse train. The critical i.e., the cut off angular frequency $p_c$ above which the CW probe is stable given by

$$p_c = 2 \left[ \frac{|W_r| r - 2 |M_r| r^2}{\beta_r''(0)} \right]^{1/2}. \tag{48}$$

The gain maximizes at the angular frequency $p_m$ which is given by $p_m = p_c / 2^{1/2}$. Positive gain, leading to MI of a CW beam, is achievable in asymmetric QW systems for a given range of angular frequencies $p$ and for specified range of values of system parameters. The parameters which govern the growth of the instability are $\beta_r''(0), |W_r|, |M_r|, p$ and normalized power $r$ of the CW probe beam. Since MI and soliton formation takes place in the same parameter space, for illustration we have used those values of $\beta_r''(0), |W_r|$ and $, |M_r|$ for which the existence of bright solitons has been predicted. Figure 3 demonstrates the MI gain spectrum $g(p)$ as a function of perturbed angular frequency $p$. The left panel is due to Kerr nonlinearity only whereas the right panel is due to both Kerr and quintic nonlinearities. Therefore, it is evident that the influence of quintic nonlinearity is to suppress the instability.

Gain of the instability maximizes at a particular frequency at given power. Figure 4 demonstrates the variation of $g_{max}$ with the power of the probe beam. When only Kerr nonlinearity is considered, the maximum gain increases slowly with the increase in power. However, if the role of quintic nonlinearity is taken into account, then $g_{max}$ initially increases slowly with the increase in power of the pump beam, then starts decreasing and finally $g_{max}$ disappears indicating that the system becomes stable against modulation instability at higher power. The quintic nonlinearity also plays significant role in determining the angular frequency $p_{max}$ of the maximum instability. In order to examine this role, we have demonstrated the variation of $p_{max}$ with power of the continuous wave in figure 5. From figure it is quite evident that $p_{max}$ increases with power and then gradually decreases to zero when both Kerr and quintic nonlinearities are present.

It should be borne in mind that the results of above analysis are also applicable to quasi-continuous/long pulse as long as the pulse duration is much longer than the period ($T_m$) of modulation instability. A quasi-continuous/ long probe pulse will be converted into a periodic pulse train of period ($T_m$) which depends on the type of nonlinearity and power of the optical field and inversely proportional to the frequency ($f_m = \frac{p_{max}}{2\pi}$) at which maximum gain occurs. For example, when Kerr nonlinearity alone is present, a probe beam of power $50 mW$ gives $f_m \sim 5.4$ GHz and $T_m \sim 0.18 ns$. Therefore, a probe pulse of duration much longer than 0.18ns is susceptible to modulation instability if power is above 50 mW.

## V. CONCLUSION:

In conclusion, we have shown the possibility of generation of ultraslow bright optical solitons due to Kerr and quintic nonlinearities in asymmetric three-coupled quantum well systems. These bright solitons arise due to nonlinearities generated by a probe pulse and two controlling laser beams. With the help of numerical simulation of nonlinear Schrödinger equation, we have demonstrated that these solitons are stable. The modulation instability of a continuous wave probe beam has been also investigated and the role of quintic nonlinearity in suppressing this instability in the same system addressed. Initially, the maximum gain of the instability increases slowly with the increase in power of the probe, then starts decreasing and finally disappears indicating that the system becomes stable against modulation instability at higher power due to quintic nonlinearity. The quintic nonlinearity also plays significant role in determining the frequency at which the gain is maximum at a given power.

## ACKNOWLEDGEMENTS:

This work is supported by the University Grants Commission, Bahadur Shah Zafar Marg, New Delhi-110001 India, through a post doctoral fellowship [Letter No. F.30-1/2009(SA-II)] and the support is acknowledged with thanks by S. Shwetanshumala. Authors would like to thank Prof. Ajoy Chakraborty, Vice Chancellor, Birla Institute of Technology, for encouragement and moral support.

FIGURE CAPTION:

1. Conduction band energy level diagram for a single period of the three-coupled asymmetric quantum well structure. Excitation scheme is shown by arrows.

2. Stable soliton shape of the probe pulse $\left|\tilde{\Omega}_P(Z,\tau)\right|$ as a function of $Z$ and $\tau$. This soliton shape has been obtained by numerically solving equation (29). In the numerical simulation the soliton solution (40) has been taken as initial condition.

3. Variation of the gain of modulation instability with instability frequency and normalized power of the probe beam.

4. Variation of maximum growth $g_{max}$ of the modulation instability with normalized power of the probe beam.

5. Variation of angular frequency $p_{max}$ of the maximum instability with normalized power of the probe beam.

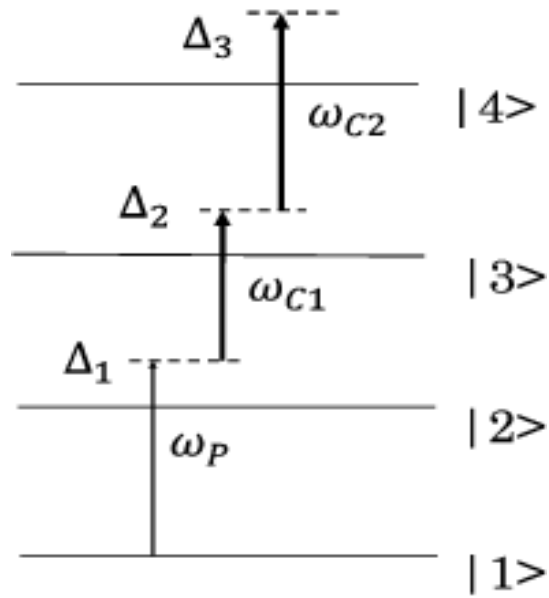

Figure 1.

Conduction band energy level diagram for a single period of the three-coupled asymmetric quantum well structure. Excitation scheme is shown by arrows.

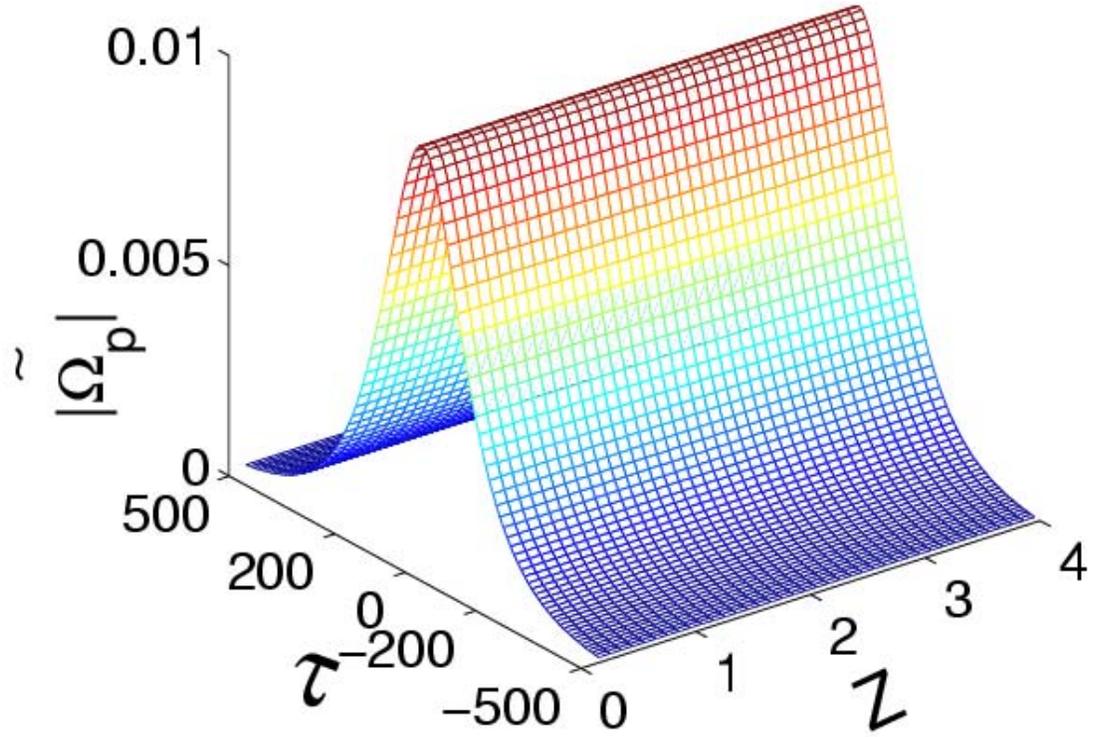

Figure 2:

Stable soliton shape of the probe pulse $|\tilde{\Omega}_P(Z,\tau)|$ as a function of $Z$ and $\tau$. This soliton shape has been obtained by numerically solving equation(29). In the numerical simulation the soliton solution (40) has been taken as initial condition.

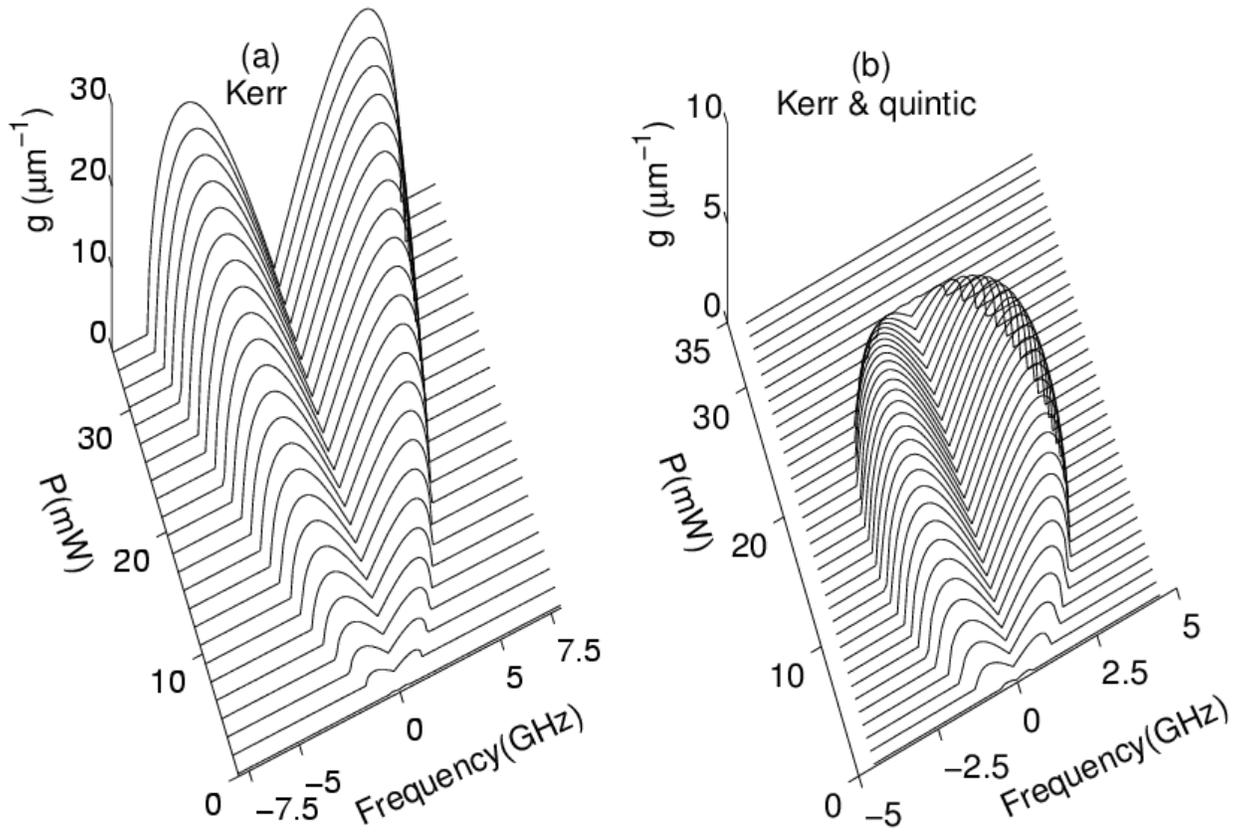

Figure 3

Variation of the gain of modulation instability with instability frequency and normalized power of the probe beam.

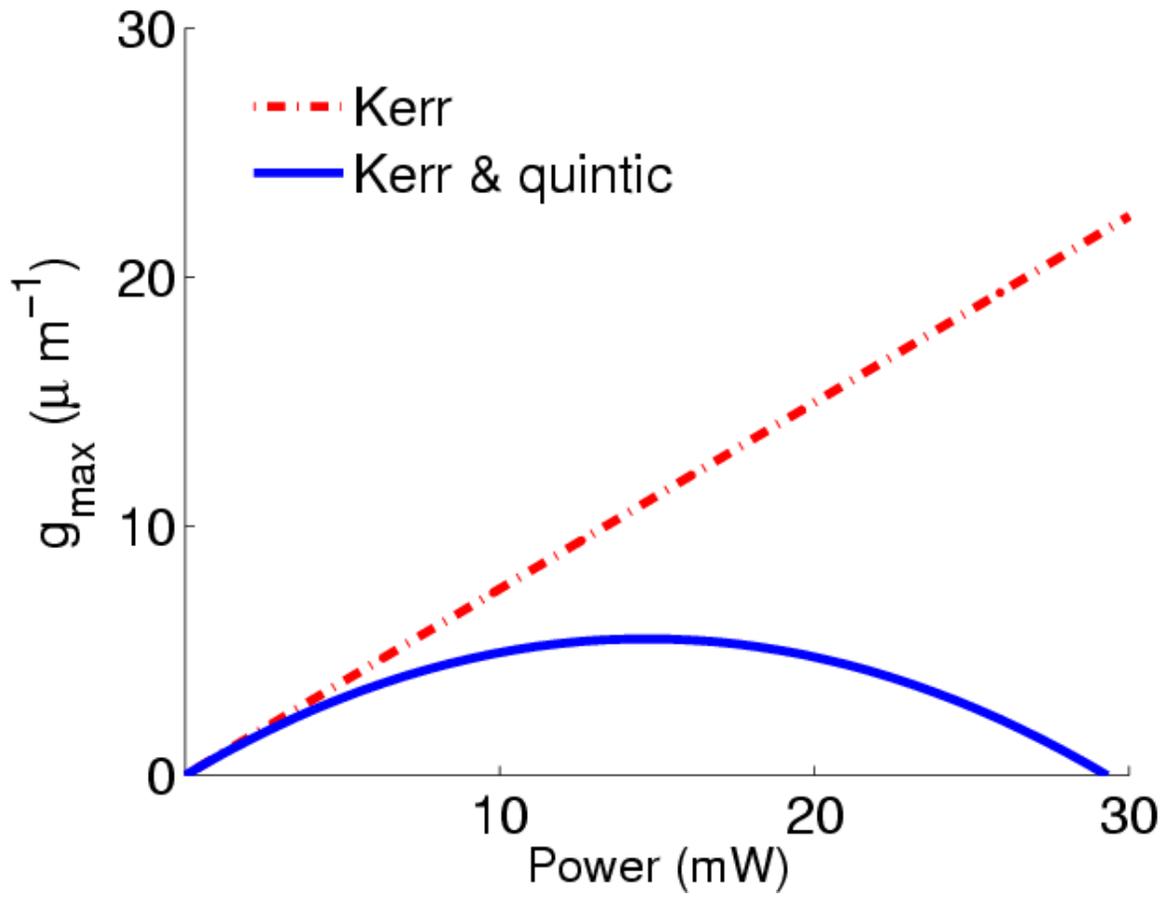

Figure 4

Variation of maximum growth $g_{max}$ of the modulation instability with normalized power of the probe beam.

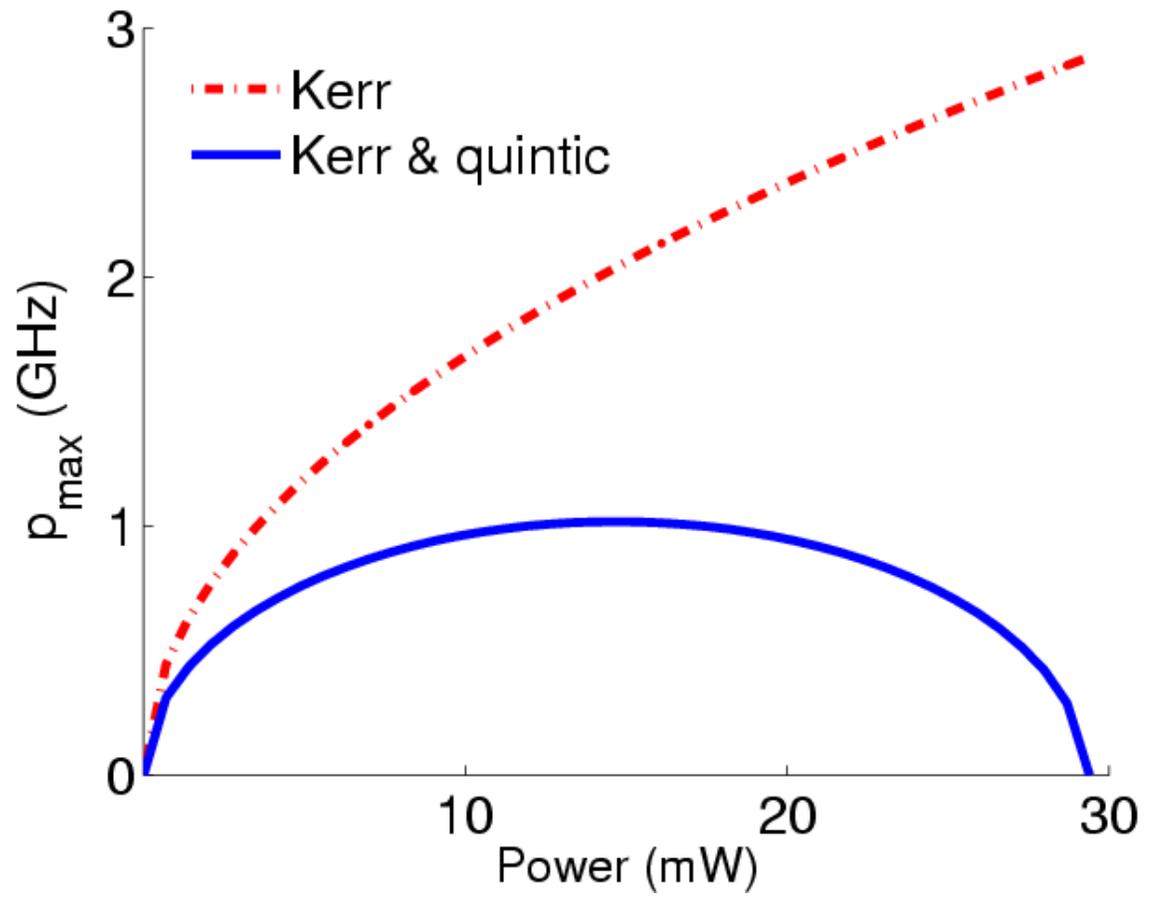

Figure 5:

Variation of angular frequency $p_{max}$ of the maximum instability with normalized power of the probe beam.